%% file: mhw.tex
\documentclass[3p,times,twocolumn,authoryear,12pt]{elsarticle}

\usepackage{caption}

\usepackage{natbib}

\usepackage{
	etoolbox, subfig, placeins, graphicx, tabularx, tikz,
	amssymb, amsmath, amsthm, 
	booktabs, hyperref
	}

\let\bbordermatrix\bordermatrix
\patchcmd{\bbordermatrix}{8.75}{4.75}{}{}
\patchcmd{\bbordermatrix}{\left(}{\left[}{}{}
\patchcmd{\bbordermatrix}{\right)}{\right]}{}{}

\newcommand{\bl}[1]{{\mathbf #1}}

\PassOptionsToPackage{hyphens}{url}\usepackage{hyperref}
\usepackage{array}
\newcolumntype{R}[1]{>{\raggedleft\let\newline\\\arraybackslash\hspace{0pt}}m{#1}}

\definecolor{green1}{RGB}{229,245,224}
\definecolor{green2}{RGB}{161,217,155}
\definecolor{green3}{RGB}{49,163,84}

\definecolor{blue1}{RGB}{222,235,247}
\definecolor{blue2}{RGB}{158,202,225}
\definecolor{blue3}{RGB}{49,130,189}

\begin{document}
\begin{frontmatter}

\title{Relax, Tensors Are Here \\ Dependencies in International Processes\tnoteref{t1}}

\tnotetext[t1]{This research was partially supported by 
the Office of Naval Research
(via grant  N00014-12-C-0066) and in part by NSF Award 1259190.}

\author[duke]{Shahryar Minhas}
\author[UW]{Peter D. Hoff}
\author[duke]{Michael D. Ward\corref{cor1}}
\ead{michael.d.ward@duke.edu}
\address[duke]{Department of Political Science, Duke University, Durham, NC 27701, USA}
\address[UW]{Departments of Biostatistics \& Statistics, University of Washington, Seattle, WA, USA}

\cortext[cor1]{Corresponding author}

\begin{abstract}
Previous models of international conflict have suffered two shortfalls. They tended not to embody dynamic changes, focusing rather on static slices of behavior over time. These models have also been empirically evaluated in ways that assumed the independence of each country, when in reality they are searching for the interdependence among all countries. We illustrate a solution to these two hurdles and evaluate this new, dynamic, network based approach to the dependencies among the ebb and flow of daily international interactions using a newly developed, and openly available, database of events among nations. 

\end{abstract}


\end{frontmatter}

\section{Motivation}

\cite{gartzke:westerwinter:2014} provide a recent paragon of research design in the field of international relations. In this research, they examine how expected changes in the distribution of power affects incentives for the so-called rising powers to sculpt their policies, so that they are more aggressive when they expect to be more powerful. At the same time, their argument states that countries have an increased incentive to undertake conflict now if they are currently more dominant than they can expect to be in the future. This notion leads to an expectation of preventative war as a consequence of power transition \citep{levy:1987}. Three hypotheses are generated: (a) shifts in military power among similar states are less likely to generate onsets of international conflict than shifts among dissimilar states, (b) military shifts among those countries with direct diplomatic relationships are not threatening and do not lead to the onset of international conflicts, and (c) changes among balances in military and economic power in states that are central in the international diplomatic network are more pertinent for war and peace than are the relative characteristics of non-central states.

The research design for the \citep{gartzke:westerwinter:2014} study is canonical by current standards in 2015. It takes a population of state dyads -- pairs of countries, such as the United States and Canada -- over the period from 1950 until 2000. This results in approximately $1.5 \times 10^6$ dyadic observations, though often they are treated as though the relationship in each dyad is symmetric, which reduces this number by half. The dependent variable is binary and has a value of $1$ in each year in which an international war begins, and otherwise is $0$. This variable is non-zero only in a very small number instances, out of the one and one-half million observations. The independent variables are created such that they, also, are expressed dyadically, even if they are really characteristics of individual states. Specifically, military capabilities are observed in each country and the difference in military capabilities within a dyad is used as a measure of military balance within the dyad.\footnote{It is actually the expected difference where expectations are generated by trend extrapolation.}  Other putatively independent variables are created in a similar way, by a comparison of the values held by each country. These include measures of voting similarity, the joint diplomatic centrality of the dyad, and others. Additionally, economic dependence is measured as the lower of the trade-to-GDP ratios of the two countries in the dyad. In this way, dyadic measures are created from the attributes of each country. This idea is longstanding in the international relations literature, perhaps stemming from the idea, which is also included in Gartzke and Westerwinter, of creating a dyadic score for democracy by taking the lower democracy score within the dyad as a characteristic of the jointness of the level of democracy. 

These data are essentially dyadic, but unconnected across the various times at which they are measured. Thus, the 1950 US$\rightarrow$Canada dyad is not linked in any way to the same dyad measured in 1951 or 1991. The dyads are also not connected to similar or even neighboring dyads. The 1950 dyad of US$\rightarrow$Canada is independent of the 1950 dyad between the US$\rightarrow$Mexico, despite sharing the US as one of the nodes. Succinctly, each dyad is assumed in empirical estimations to be independent of every other dyad. If this is a correct assumption---and it is not---then using a simple regression framework will reveal useful information about the relations among the independent and the dependent variable. Such a regression table is what is provided in the canonical example. Table~1 in the Gartzke \& Westerwinter manuscript (page 10) shows that in a logistic regression expected changes in capabilities have a positive influence on the probability of a war onset and the measure of interest similarity has a strong but negative impact on war onset. The respective coefficients are large, and their respective standard errors are small. The putative statistical significance of these coefficients is asserted to support the underlying hypotheses.\footnote{We ignore for this discussion the use of inferential statistics in a population.} 

The formal representation of this data structure is given by a time series of vectors comprising the measured dyadic relations at each point in time; for each dyad $i\rightarrow j$ and time $t$, $y_{i,j,t}$ denote the relation of the dyad at that time. The total structure has size $m\times (m-1)\times n$, where $m$ is the number of countries and $n$ is the number of time periods, and could be related to other similarly defined structures for other variables that might serve as covariates, as in the case of the democratic peace literature, or in the Gartzke and Westerwinter example, above. Such a structure doesn't necessarily preserve, but rather generally destroys, the dependencies among the nodes and across time.

It is worth noting that this basic research design is well established, and largely unchanged since the 1980s. Examples of virtually the same design -- using virtually the same data -- include \cite{maoz:abdolali:1989}, \cite{bremer:1992}, \cite{maoz:russett:1993},  \cite{bremer:cusack:1995}, and \cite{pevehouse:russett:2006}, as well as many, many others. Moreover, the same database has been utilized in most of these studies, and while it has been updated since the 1980s virtually all quantitative research on international conflict and cooperation at the dyadic level has used the same, canonical database: the Militarized Interstate Dispute data that were initially created by the Correlates of War project in the 1980s \citep{palmer:etal:2015}. Thus these data and this research design have been a constant of research over the past several decades. What is clear is that it is hard to underestimate the resilience of the standard regression framework in spite of demonstrations that it can lead to faulty conclusions in the presence of higher-order dependencies in dyadic data structures: \cite{hoff:raftery:etal:2002,hoff:ward:2004,hoff:2005,hoff:2010}. See also \cite{cranmer:desmarais:2011}, \cite{desmarais:cranmer:2012}, \cite{cranmer:desmarais:etal:2012}, and \cite{snijders:koskinen:etal:2012}. A nice summary and additional development can be found in \cite{stewart:2014}. 

Below we outline a different approach to the regression framework that embraces the dependency across actors as well as over time, permitting a regression based approach to estimating the parameters of dynamic networks of political actors. We then apply this to a new dyadic database, which is completely independent of the COW MID data, on conflict and cooperative events among countries.

\section{Modeling Approach}

Let relational data that is longitudinal be represented as a series of matrices $\{\bl Y_t : t = 1, \ldots n\}$, with each $\bl Y_t$ representing the actions or flows between the pairs of actors, in our case countries, at time $t$. That is, $\bl Y_t$ is the $m\times m$ matrix whose $(i,j)^{th}$ entry is $y_{i,j,t}$, the measured interaction between actors $i$ and $j$ at time $t$. If these data are binary -- such as onsets of war or militarized disputes -- this is a different but equivalent representation of the canonical data in the quantitative study of war, as in the case of Gartzke and Westerwinter (2014). But such binary data are also considered by a different research tradition to represent {\em networks}. One such matrix  $\bl Y_t$ is shown in Figure~\ref{fig:matrix}. Here at time $t$ country $i$ has initiated disputes with $j$ and $k$, and $k$ has initiated a dispute with $i$. 

\begin{figure}[ht]
\begin{equation*}
Y_{t} = \bbordermatrix{
	~ & i  & \ldots & j & \ldots & k \cr
	i & 0  & \ldots & 1 & \ldots & 1 \cr
	\vdots & \vdots & \ddots & \vdots & \ddots & \vdots  \cr
	j & 0  & \ldots & 0  & \ldots & 0 \cr 
	\vdots & \vdots & \ddots & \vdots & \ddots & \vdots \cr
	k & 1  & \ldots & 0  & \ldots & 0 \cr 		
	}
\end{equation*}
	\caption{Adjacency matrix representation of a dyadic, relational measure for one time point.}
	\label{fig:matrix}
\end{figure}

\citet{snijders:2001} developed an agent based approach to modeling such binary networks that considers the linkages among the actors in terms of decisions, such that utility models could be implemented to generate the linkages. This strategy does enable the specification of a variety of network dynamics that may exist in the data, but the decision logic is homogeneous, with the same utility applying to each actor. Current versions of this approach are perhaps better known as Jackson-Wohlensky networks, though the estimation of the parameters is typically given, rather than estimated.

An alternative approach has been to specify dynamic latent variables, in which each $\bl Y_t$ is modeled as a function of actor specific latent variables. The goal of this approach is to collapse information contained in the network about higher-order dependencies to a lower dimensional latent space. Embedding the network information onto this lower dimensional space has the added benefit of providing meaningful visualizations of how likely actors are to interact with each other due to network influences. This approach was first introduced in the political science literature by \citet{ward:hoff:2007} and expanded by \citet{ward:ahlquist:etal:2012}, and has been widely employed. Similar, recent approaches include \citet{durante:dunson:2014}. 

Common to each of these approaches is an effort to account for higher-order dependencies that can emerge in network data. Examples of these dependencies include concepts such as reciprocity and transitivity that are familiar to the international relations literature. Broadly, these dependencies can be placed along two dimensions: second and third-order dependencies. 

Second-order dependencies refer to what is often described as reciprocity in the context of directed relationships. Reciprocity is a not a new concept to the field of international studies; it has its roots in previous theories of cooperation and the evolution of norms between states \citep{richardson:1960,choucri:north:1972,ward:1981,cusack:ward:1981,ward:1984,goldstein:1991}. This concept has particular relevance in the conflict literature, as we would expect that if, for instance, Iran behaved aggressively towards Saudi Arabia that this would induce Saudi Arabia to behave aggressively in return. The prevalence of these types of potential interactions within directed dyadic data structures directly challenges the basic assumption of observational independence.

An example of a third-order dependency is transitivity, which follows the familiar logic of ``a friend of a friend is a friend''. The importance of these types of relationship is being increasingly recognized in the literature as well (see, for example, \citealp{lai:1995,manger:etal:2012,kinne:2013}). In binary data, transitivity describes the dependence among three actors $i$, $j$, $k$ in which $i$ and $j$ are more likely to be linked if linkages already exist between $i - k$ and $j - k$. The principal idea behind these dependencies is that knowing something about the relationships between $i-k$ and $j-k$ may reveal information about the relationship between $i-j$, even if it is not directly observed. This type of network dynamic might be particularly important for understanding the development of cooperative relationships. The consequence of not incorporating this type of dependency into our modeling framework is the same as above, and, as a result our models, will suffer from misspecification bias and will be more likely to result in type 1 errors. 

While both the agent and latent space approaches effectively incorporate statistical interdependence among the actors,  agent-based approaches assume that this effect is homogeneous among all dyads. The latent variable based approach allows heterogeneity between dyads. The downside, however, of earlier work using the latter approach is that explicit parameterizations of the effect of reciprocity and transitivity could not be calculated due to the simple structure of the latent variables. Instead all of the information contained within those higher order dependences was collapsed into a lower dimensional latent space. \citet{hoff:2014} provides a novel approach that combines both of these approaches in terms of their strengths: it has an explicit representation of the dependence, specifically, reciprocity and transitivity, between dyads, but also allows for heterogeneity among nodes.

\section{Tensor Model}

For many dyadic, relational data in international relations we do not have just a snapshot of interactions in time, but an actual time series of observations. Say that $(\mathbf y_t)$ represents a time series of vectors representing the dyadic interactions for $m$ countries through the period $t = 1, \ldots, n$, so that $\mathbf y_t \in \mathbb{R}^{m^2}$ (this is just the matrix $\bl Y_t$ rewritten as a vector). A simple way to model the evolution of this time series is as a function of past behavior. A modeling framework that lends itself to this type of set up is a vector autoregression (VAR) model. The VAR model has been employed in macroeconomic research ever since \citet{sims:1980} introduced the approach, and it remains a useful tool for political scientists, as evidenced by the work of \citet{brandt:freeman:2006}. The VAR model is a generalization of the univariate autoregressive model that accounts for linear interdependencies among multiple time series. The benefit of this approach is that it models a set of time series as resulting from a dynamic and endogenous process, rather than imposing a priori assumptions of exogeneity between the various explanatory variables. 

A first-order VAR model posits that:
\begin{eqnarray}
	\bl y_t &=& { \bl \Theta \bl y_{t-1} +\bl e_t}\\
	E[\bl e_t] &=& 0 \\
	E[\bl e_t \bl e^T_s] &=& \begin{cases}  \bl \Sigma &\mbox{if } t=s, \\ 0 &\mbox{otherwise}\end{cases}
\end{eqnarray}

\noindent where $\bl y_t$ has been demeaned to eliminate intercepts such that $\sum_t y_{i,j,t}/n \equiv 0$. This is easy to estimate, but has extreme data requirements since $\bl \Theta$ has $m^4$ entries. One basic solution to this is to use a bilinear model such that the regression matrix is given by $\bl \Theta = \bl B \otimes \bl A$, where $\otimes$ is the Kronecker product. As a result of this specification we obtain a basic specification that looks relatively familiar: $\bl Y_t = \bl A\bl Y_{t-1}\bl B^T$. 

In terms of this formulation we have, for each $i$ and $j$, $E[y_{i,j,t}| \bl Y_{t-1}] = \sum_{i'} \sum_{j'} a_{i,i'} b_{j,j'} y_{i',j',t-1}$, which makes clear that the {\em regression parameters are multiplicative functions} of the basic bilinear parameters. The $a_{i,i'}$ capture how previous actions of $i'$ affect $i$ and $b_{j,j'}$ shows how actions that target $j$ are influenced by prior actions toward $j'$.

Given that we are interested in estimating the endogenous relationships between how a set of network variables about the interactions of states evolve over time, it will be useful to quickly go over how the data we are modeling is structured. Recall that in Figure~\ref{fig:matrix}, the matrix contains a snapshot of interactions on a particular dimension occurring between a set of $m$ actors at a given point in time. Now we are moving to a setting through which we can estimate how a network evolves over time, thus we have a time series of matrices $\{{\bl Y}_t : t = 1, \ldots, n\}$. Additionally, we want to estimate a model in which we are not required to make a priori assumptions about exogeneity among our covariates, so we employ a modeling framework that allows us to model the endogenous effects of a set of $v$ relational covariates over time. 

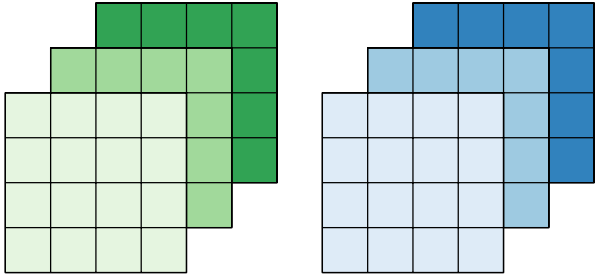
\begin{figure}[ht]
	\centering
	\resizebox{.5\textwidth}{!}{\input{tensorViz.tex}}
	\caption{Tensor representation of longitudinal dyadic, representational measures. The green and blue colors represent different relational measures and darker shading indicates later time periods. Specifically, we show a tensor with dimensions of $4 \times 4 \times 2 \times 3$, where 4 represents the number of actors, 2 the number of relational measures, and 3 the number of time points.}
	\label{fig:tensViz}
\end{figure}

Thus the data that we are modeling no longer takes the shape of a simple series of two dimensional matrices, but instead should be thought of as a multiway array or tensor, $\bl Y$, of dimension $m \times m \times v \times n$, where $m$ corresponds to the number of countries, $v$ to the number of variables, and $n$ to the number of time periods.  Figure~\ref{fig:tensViz} provides an example of this for the simple case of four countries, three time points, and two relational measures. 

Now take $\bl Y_t$ to be the time $t$ component of $\bl Y$ (so now, instead of an $m\times m$ matrix, a tensor incorporating each of the $v$ variables). The regression problem then becomes one of regressing the relational tensor $\bl Y_t$ from time $t$ on the tensor $\bl X_t =\bl Y_{t-1}$ from time $t-1$ in a parsimonious way. \citet{hoff:2014} proposes and develops the following multilinear generalization of the bilinear regression model: 

\begin{equation}  
	\bl Y = \bl X \times \{ \bl B1_{ii'}, \bl B2_{ii'}, \bl B_3\} + \bl E ,  
	\label{eqn:mltr}
\end{equation}

\noindent where ``$\times$'' is a multilinear operator known as the ``Tucker product''. Here $\bl Y$ and $\bl X$ are $m \times m \times v \times n$ tensors and each entry for $\bl X$ is lagged by one period. 

In this model $\bl B1_{ii'}$ and $\bl B2_{ii'}$ are $m \times m$ matrices of regressions coefficients, and are similar to the parameters $\bl A$ and $\bl B$, respectively, discussed above. The key difference, however, is that we are no longer running separate models for every $v$; rather we are leveraging each of the $v$ dyadic, relational variables to help predict the others. Thus $\bl B1_{ii'}$ and $\bl B2_{ii'}$ represent how previous actions along any of the $v$ parameters affects future interactions. For example, the $l j$ coefficient in $\bl B1_{ii'}$ measures the effect of an interaction from actor $j$ to actor $k$ along any of the $v$ parameters on the likelihood of an interaction from $l$ to $k$, in any of the parameters, in the next time period.

$\bl B_3$ is a $v$ by $v$ matrix of coefficients representing the direct effects of the lagged dyadic, relational parameters on one another. In other words, the $u w$ coefficient of $\bl B_3$ provides a measure of the effect of the value of parameter $w$ in a dyad on the value of parameter $u$ in that dyad during the next time period. In order to incorporate explicit representations of higher-order dependencies we extend the $\bl X$ tensor along its third dimension. 

First, we add in reciprocity, which captures the tendency for actions to be reciprocated from one node, $i$, to another, $i'$ for a variable $w$. More explicitly, if $y_{i,i',w,t}$ (that is, the $i$ to $i'$ relationship in variable $w$ at time $t$) is large, and reciprocal relationships exist, we would expect $y_{i',i,w,t+1}$ to be large as well. In the formulation of $\bl X$ we have presented so far, its third dimension has length equal to $v$; we add $v$ additional slices to the tensor along its third mode to incorporate reciprocity, so that it now has dimension $m \times m \times (2*v) \times n$. The first $v$ slices across this third mode are the same as before, $y_{i,i',w,t-1}$, but the last $v$ slices are defined using the reciprocal lagged elements, $y_{i',i,w,t-1}$. The general form of our tensor regression remains the same as what was shown in Equation~\ref{eqn:mltr}, but now $\bl B_3$ is a $v$ by $(2*v)$ matrix, in which the last $v$ columns capture the tendencies of actions to be reciprocated by other actions at the next time point.

Next, we add in a measure of transitivity to account for the possibility of third-order dependencies. An implication of transitivity is that existing linkages $y_{i,i'',w,t}$ and $y_{i',i'',w,t}$, imply that $y_{i,i',w,t+1}$ may form a linkage as well. To incorporate this additional term, we again add $v$ slices to the third mode of our $\bl X$ tensor. The elements of these last $v$ slices are  $\sum_{i''} ( y_{i,i'',w,t} + y_{i'',i,w,t}) ( y_{i',i'',w,t} + y_{i'',i',w,t})$. Similar to when reciprocity was incorporated, $\bl B_3$ now expands to a $v$ by $(3*v)$ matrix, in which the last $v$ columns capture how relations that $i$ and $i'$ share with common targets leads to future interactions between $i$ and $i''$.

The parameters in our completed model formulation have the following dimensions: 

\begin{itemize}
	\item $\bl Y$: $m \times m \times v \times n$, 
	\item $\bl X$: $m \times m \times (3*v) \times n$, 	
	\item $\bl B1_{ii'}$ and $\bl B2_{ii'}$: $m \times m$, 	
	\item $\bl B3$: $v \times (3*v)$,
\end{itemize}

\noindent where $m$ refers to the number of countries, $v$ the number of relational parameters, and $n$ the number of time periods. This framework can be referred to as a relational multilinear regression. Like previous work using latent spaces to incorporate network characteristics, we are able to model actor heterogeneity in the network, however, this framework now also enables the explicit measurement of higher-order dependencies. \citet{hoff:2014} provides a Bayesian implementation of this model via Gibbs sampling using semi-conjugate prior distributions to simulate parameter values from the corresponding posterior distribution.\footnote{Code for this Gibbs sampler can be found here: \url{http://www.stat.washington.edu/people/pdhoff/Code/hoff_2014/}.}

\section{ICEWS Dyadic Event Database}

We use this modeling framework to estimate models for verbal and material cooperation and conflict between states. Information on these dimensions is collected from the Integrated Crisis Early Warning System (ICEWS) event data project. The ICEWS data along with extensive documentation are available on \url{dataverse.org} \citep{icews:2015:aggregations,icews:2015:data}. The ICEWS event data are constructed by applying natural language processing techniques based on natural language processing and graph theory \citep{boschee:natarajan:etal:2013} to a corpus of about 30 million media reports from about 275 local and global news sources in or translated to English.  Each media report is coded in accordance with an ontology of events that is derived from the Conflict and Median Event Observation (CAMEO) ontology  \url{http://eventdata.psu.edu/cameo.dir/CAMEO.CDB.09b5.pdf}. 

This results in relevant reports being tagged as involving a set of actors and by the interactions that took place between them. We choose to focus on interactions among the top fifty countries by GDP during time period from the beginning of 2001 to the end of 2014. And, like all good social scientists, we use a $2 \times 2$ table to classify; herein, across the distinctions between actions and words on the one hand and conflict and cooperation on the other.  Thus, we aggregate events for this set of countries and temporal span into the so-called ``quad variables''. \citet{yonamine:2011} provides a definition for each:

\begin{itemize}
	\item \emph{Verbal Cooperation}: The occurrence of dialogue-based meetings (e.g. negotiations, peace talks), statements that express a desire to cooperate or appeal for assistance (other than material aid) from other actors.
	\item \emph{Material Cooperation}: Physical acts of collaboration or assistance, including receiving or sending aid, reducing bans and sentencing, etc.
	\item \emph{Verbal Conflict}: A spoken criticism, threat, or accusation, often related to past or future potential acts of material conflict.
	\item \emph{Material Conflict}: Physical acts of a conflictual nature, including armed attacks, destruction of property, assassination, etc.
\end{itemize}

Given that the quad variables are generated using an event data approach, we have a very low level of temporal aggregation available to us through which to measure directed interactions between countries. In this analysis, we focus on the monthly level of aggregation, thus leaving us with a large time series of directed, dyadic, relational observations for each quad variable. 

\begin{figure*}[ht]
	\centering
	\includegraphics[width=1\textwidth]{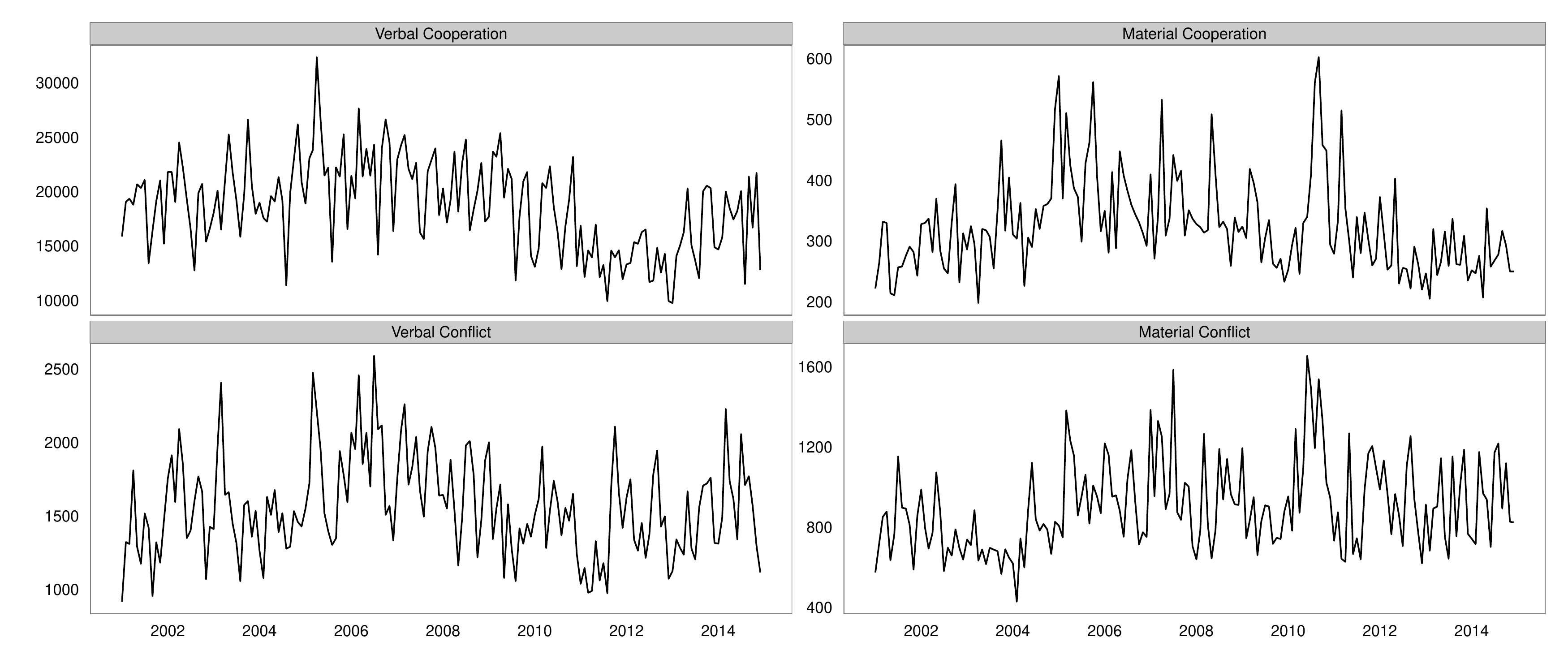}
	\caption{Quad variable counts over time.}
	\label{fig:dvAgg}
\end{figure*}

\begin{figure*}[ht]
	\centering
	\includegraphics[width=1\textwidth]{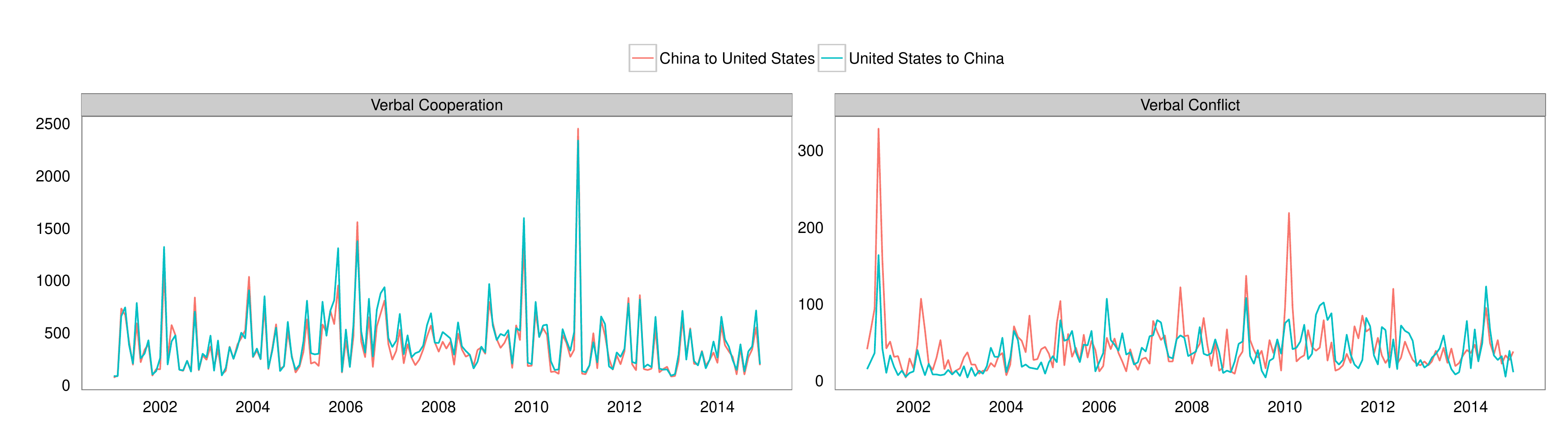}
	\caption{Quad variable counts over time for interactions between the United States and China.}
	\label{fig:dvParticular}
\end{figure*}

In Figure~\ref{fig:dvAgg}, we show the total number of event counts falling under each of these variables at the monthly level. Looking across the panels we can see that verbal cooperation events between countries is much more frequent than the other relational variables we use here. Further, the verbal event variables, in general, are more frequently occurring than the material events. This itself may simply reflect the fact that verbal actions are less costly to engage in than material ones, thus they occur more frequently. Despite this the number of material conflictual events, that is, events involving a country engaging in activities such as imposing a blockade, are not rare. For many months during the time period shown, the number of material conflictual events involving the 50 countries in our analysis exceeded a 1,000. Material cooperation, on the other hand, is a much less frequent event within the context of our sample. This is likely due to the fact that by design we chose the fifty countries with the highest average GDPs during the period of 2001 to 2014, and these countries are less likely to provide material aid to each other than they would be to poorer countries. 

An important benefit to using these relational measures is that they display interesting variation over time. Most measures used in the extant literature to estimate the dyadic relations between states exhibit relatively little temporal variation (e.g., MIDs), and, if they do, they are often not available at low levels of temporal aggregation without substantial loss of country coverage. The quad variables shown here suffer from neither of those drawbacks. Thus the addition of these variables to the set of databases used in political science research may enable the field to develop a richer understanding of how directed relationships between states evolve. 

An example of how we can trace interesting features of state interaction over time using these variables is shown in Figure~\ref{fig:dvParticular}. Here we visualize the directed relations between the United States and China for the two verbal quad variables at the monthly level across our period of analysis. The blue line in the left panel illustrates verbal cooperative actions that the United States sent to China, while the red line designates the same type of actions that China sent to the United States. 

In the case of verbal cooperation, we can see that the two lines almost completely overlap with one another, indicating that many of these events either occur in tandem or are reciprocated almost immediately. The panel on verbal conflict shows similar patterns of reciprocity as verbal cooperation, but unlike cooperative interactions between the United States and China verbal conflictual events are usually driven to a greater degree by one party or another. For example, in February of 2010, we can see a peak in the number of verbal conflictual events sent by China to the United States. This likely was related to the announcement of an arms sale package to Taiwan by the Obama administration. 

To begin to model the processes driving the quad variables we turn to the the relational multilinear regression framework introduced earlier. We run two separate models. In the first, we group together the verbal and material conflict variables, and, in the second, the verbal and material cooperation variables. We model the processes of cooperation and conflict separately in this iteration of the project to ease the interpretability of the parameter estimates. Additionally, if we run a tensor model using all four of the covariates, as opposed to running two separate tensor models as suggested above, the model performance noticeably declines. 

For both models then, this leaves us with a $\bl Y$ tensor that has dimensions $50 \times 50 \times 2 \times 167$, where 50 corresponds to the number of countries, 2 to the number of relational variables, and 167 to the number of months included in our analysis. Given that we are including explicit representations of higher-order dependencies, our $\bl X$ tensor has dimensions $50 \times 50 \times 6 \times 167$. We also apply a normal quantile-quantile transformation to the time-series for every sender-target-variable type triplet and further centered and scaled each, so that their empirical distributions are approximately standard normal. 

\section{Results}

\begin{figure}[ht]
	\centering
			\includegraphics[width=.45\textwidth]{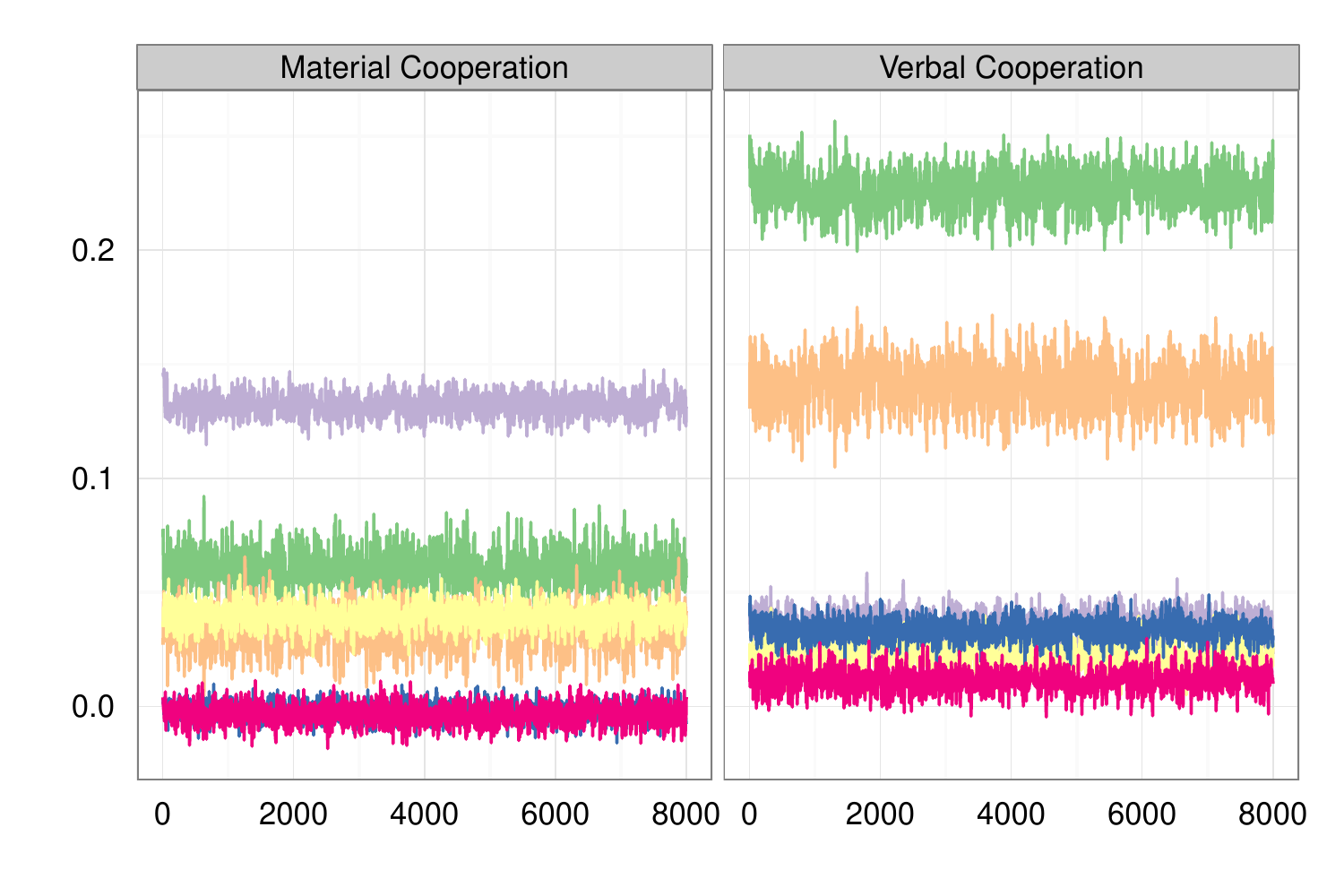}
	\caption{Trace plots for each of the six $\bl B_3$ parameter estimates from the cooperation model. Different colors correspond to different parameter estimates.}
	\label{fig:trace}
\end{figure}

Figure~\ref{fig:trace} shows the trace plots after running 8,000 iterations of the Gibbs sampler for each of the six $\bl B_3$ parameter estimates from the cooperation model. Trace plots for the conflict model produce results that are similar, in terms of indicating that the iterative
estimation algorithm quickly converged. The starting value for these Gibbs samplers were the least square estimates and we can see that each parameter converges to its stationary distribution almost immediately. We chose to use the first 1,000 iterations as burn-in, leaving us with 7,000 iterations for every parameter through which to conduct posterior analysis.

\subsection{Direct, Reciprocal, \& Transitive Effects}

\begin{figure*}[ht]
	\centering
	\begin{tabular}{cc}
		\subfloat[][Cooperation Model]{
			\includegraphics[width=.45\textwidth]{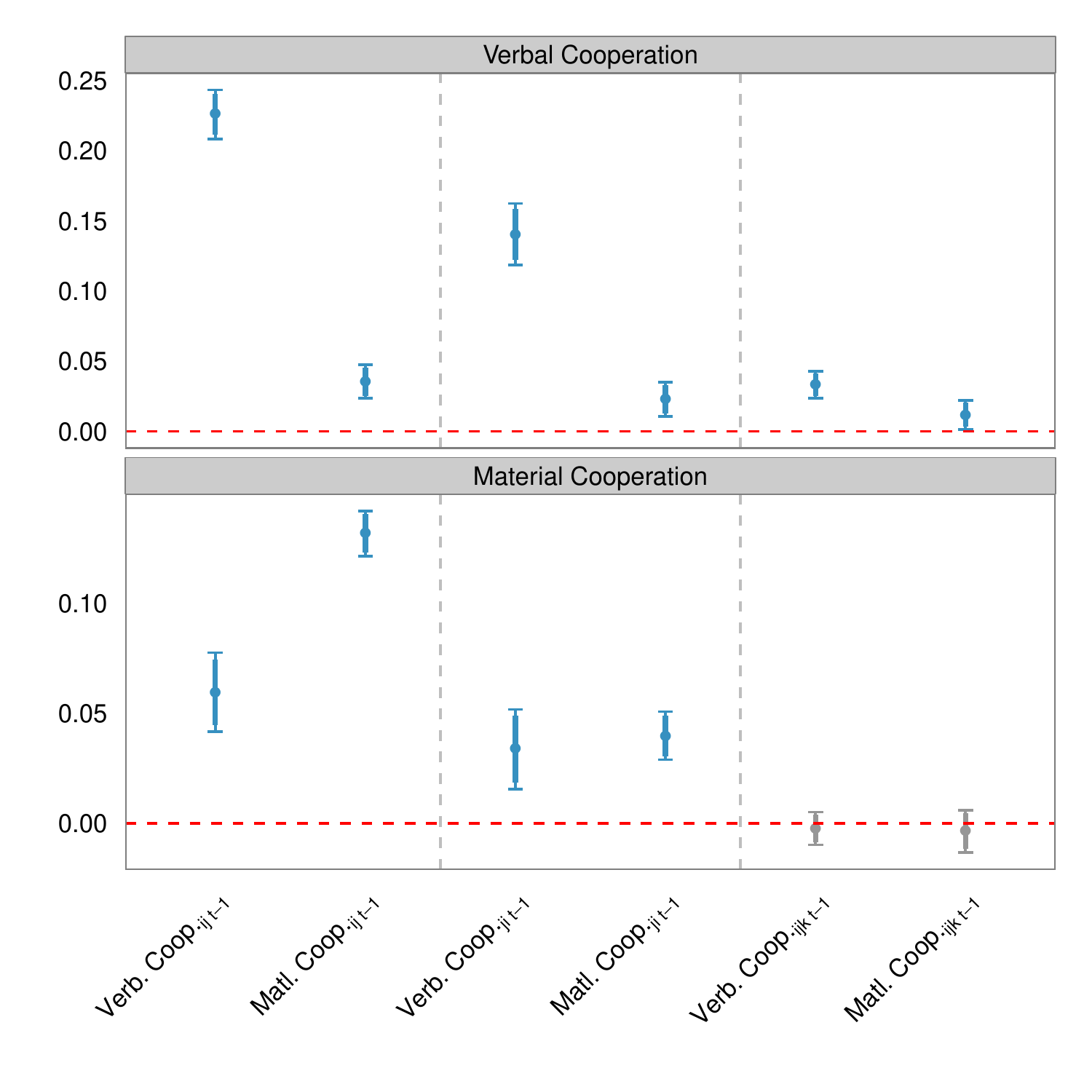}
		\label{fig:coopB3}}	&	
		\subfloat[][Conflict Model]{
			\includegraphics[width=.45\textwidth]{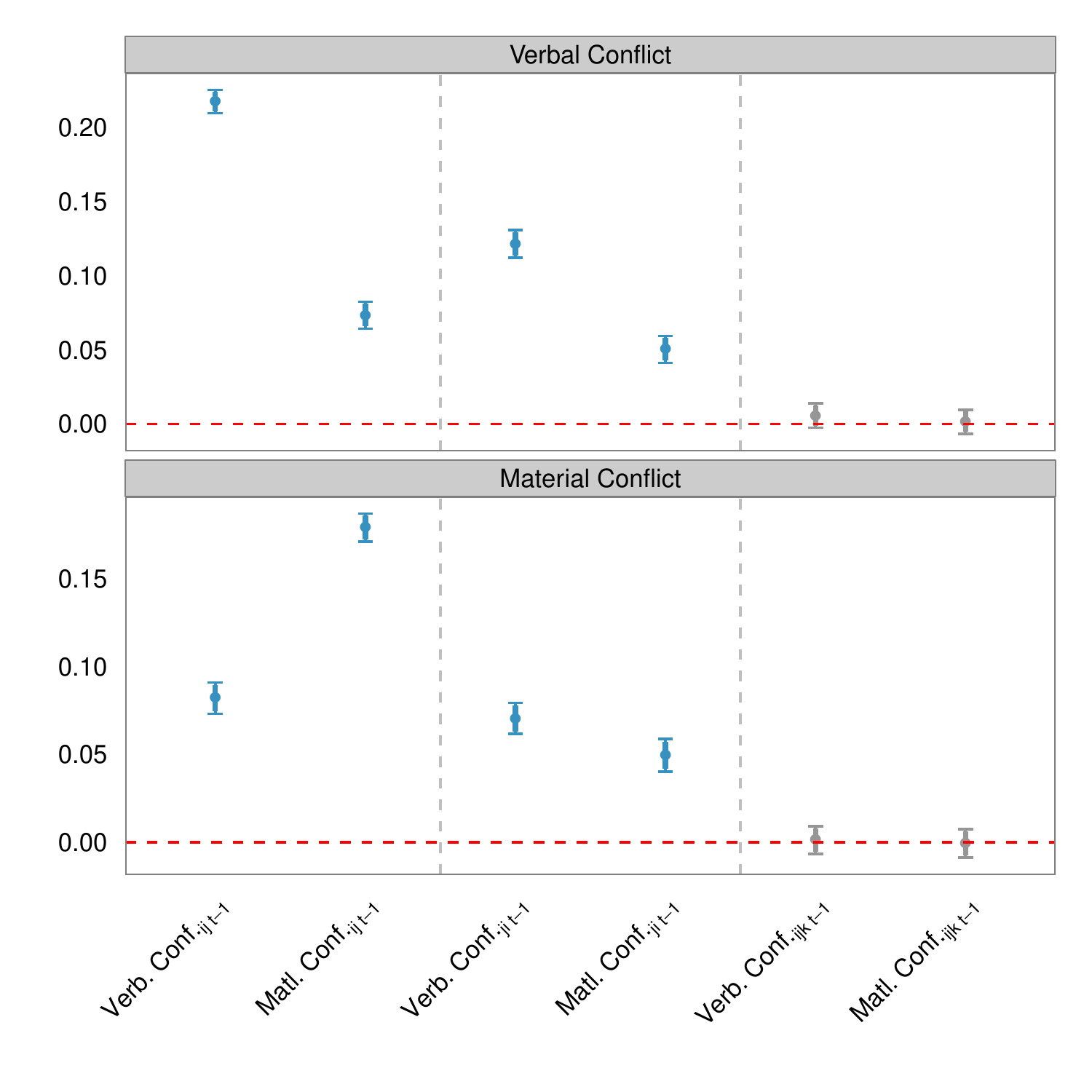}
		\label{fig:confB3}}
	\end{tabular}
	\caption{Relational multilinear regression results for $\bl B_3$. Darker colors indicate that the coefficient estimate is significantly different from zero at a 95\% credible interval, while lighter the same for a 90\%. Grey indicates that the estimate is not significantly different from zero at either of those intervals.}
	\label{fig:B3}
\end{figure*}

The posterior distributions of the $\bl B_3$ coefficients are summarized in Figure~\ref{fig:B3}. The panel on the left corresponds to the tensor model on cooperation and the right to conflict. Also note that though we split up the verbal and material cooperation dependent variables in Figure~\ref{fig:coopB3}, they are estimated simultaneously using the modeling framework discussed earlier -- the same holds for the conflict model. Additionally, given the transformations we made to the time series for each of our parameters, the relative importance of our parameters can be assessed by comparing their coefficient estimates.

Along the x-axis of this figure, we list the direct ($ij$), reciprocal ($ji$), and transitive lagged estimates ($ijk$) for each variable. The direct estimate represents the effect of a particular variable from $i$ to $j$ on the future value of that variable from $i$ to $j$. The reciprocal estimate designates the effect of actions from $j$ to $i$ on future actions from $i$ to $j$. Last, the transitive estimate assesses the effect on action between $i$ and $k$ in the future given that $i$ and $j$ and $j$ and $k$ are already interacting with one another.

For each of the dyadic, relational variables examined in Figure~\ref{fig:B3}, the largest estimated coefficient comes from the direct, lagged parameter, indicating that the strongest predictor of relations at a given point a time are what they were in the period before. For instance, the regression estimates for the Material Conflict variable, shown in the bottom half of Figure~\ref{fig:confB3}, the parameter estimate for Material Conflict$_{ij, t-1}$ is twice as large as that of Verbal Conflict$_{ij, t-1}$. This general trend persists for the predictors of Verbal Conflict and for the results of the cooperation model. 

The lagged effect, however, is not the only powerful and precise predictor. For both the cooperation and conflict models, the effect of both of the direct, lagged variables is precisely estimated and positive. Thus, for example, in the case of our model on cooperation, we find that higher levels of verbal cooperation between a pair of countries is also likely to lead to higher levels of material cooperation and vice versa. The same finding holds for the parameters measuring the direct effects in the conflict model.

The results discussed so far correspond to the type of parameters that existing latent space approaches can already estimate. But the novel component of this new framework comes with the ability to estimate higher-order effects for each of these covariates. Thus we turn to the set of $ji$ parameter estimates in the panels shown in Figure~\ref{fig:B3}. In every case, we find that the reciprocal effects are precisely estimated for both lagged parameters, and that there are a number of instances in which the reciprocal effects are of similar magnitude to the direct effects. The implication of this, though potentially not surprising, is that countries tend to reciprocate behavior. For instance, in the case of our cooperation model, we find that material cooperation in a $ji$ pair is likely to lead to reciprocal material cooperation in the $ij$ pair in the next period. More interestingly, we also find evidence in both our models that verbal actions are reciprocated by material ones and vice versa. 

The key implication of these findings for the reciprocal parameters in our models is that second-order dependencies have a significant role to play in our understanding of interactions between states. As noted earlier, ignoring these types of dependencies not only closes off interesting theoretical avenues of analysis, but also can lead to inferential mistakes. The last set of coefficient estimates contained in $\bl B_3$  are the transitive effects of each of the lagged covariates. We find less support for this particular type of higher-order dependency in our analysis. In the conflict models, none of the parameters used to estimate transitive effects have an indentifiable role in explaining verbal or material conflict. In the case of our cooperation model, we do find some evidence that transitivity comes into play in explaining verbal cooperation. Though the effect is small, the implication is that countries are likely to engage in verbal cooperation with countries to whom they they share indirect verbal or material ties already. The same effect, however, does not hold for material cooperation, likely because providing material support is more costly. 

This all makes a good deal of intuitive sense: if $i$ and $j$ are at a high level of conflict on issue A, and $i$ and $k$ are at war on issue B, why would we expect $j$ and $k$ to enter a conflict with one another? On the cooperative side of things, it seems more plausible. And this is exactly what the empirical estimates suggest.

\begin{figure*}[ht]
	\centering
	\begin{tabular}{c}
		\subfloat[][Cooperation Model]{
			\includegraphics[width=1\textwidth]{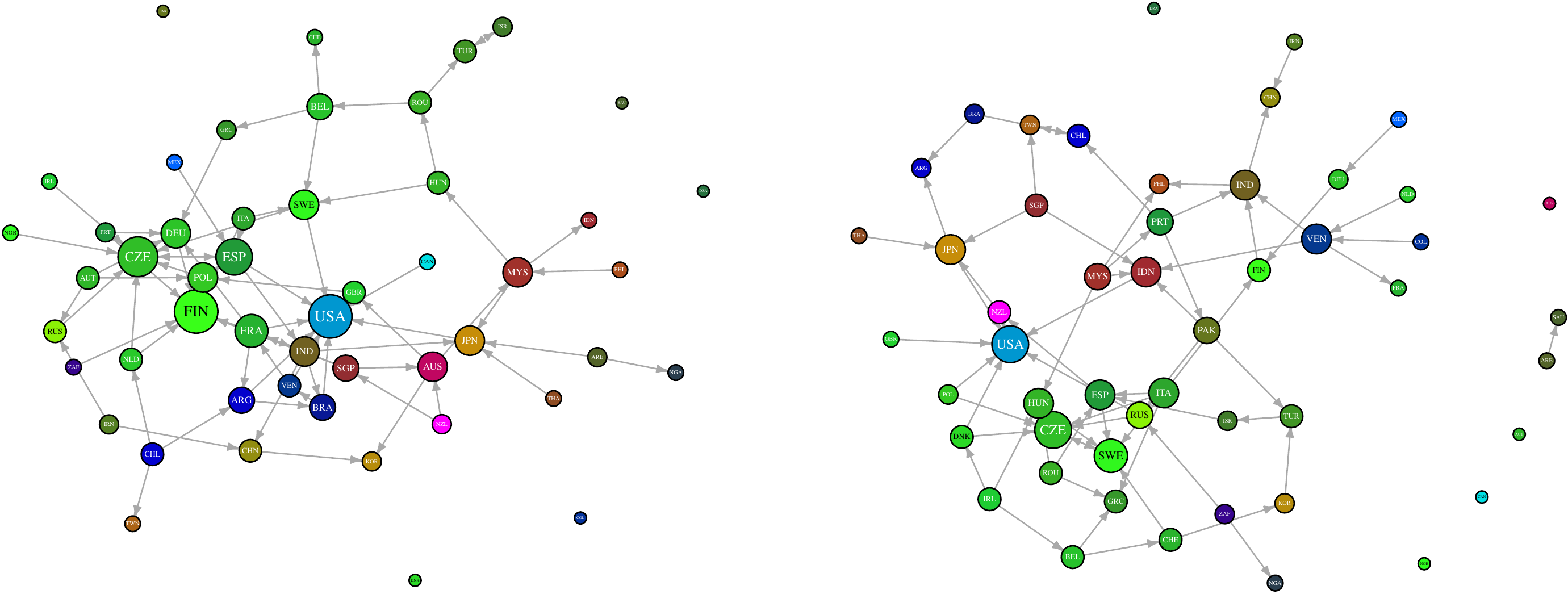}
		\label{fig:coopB12}} \\	
		\subfloat[][Conflict Model]{	
			\includegraphics[width=1\textwidth]{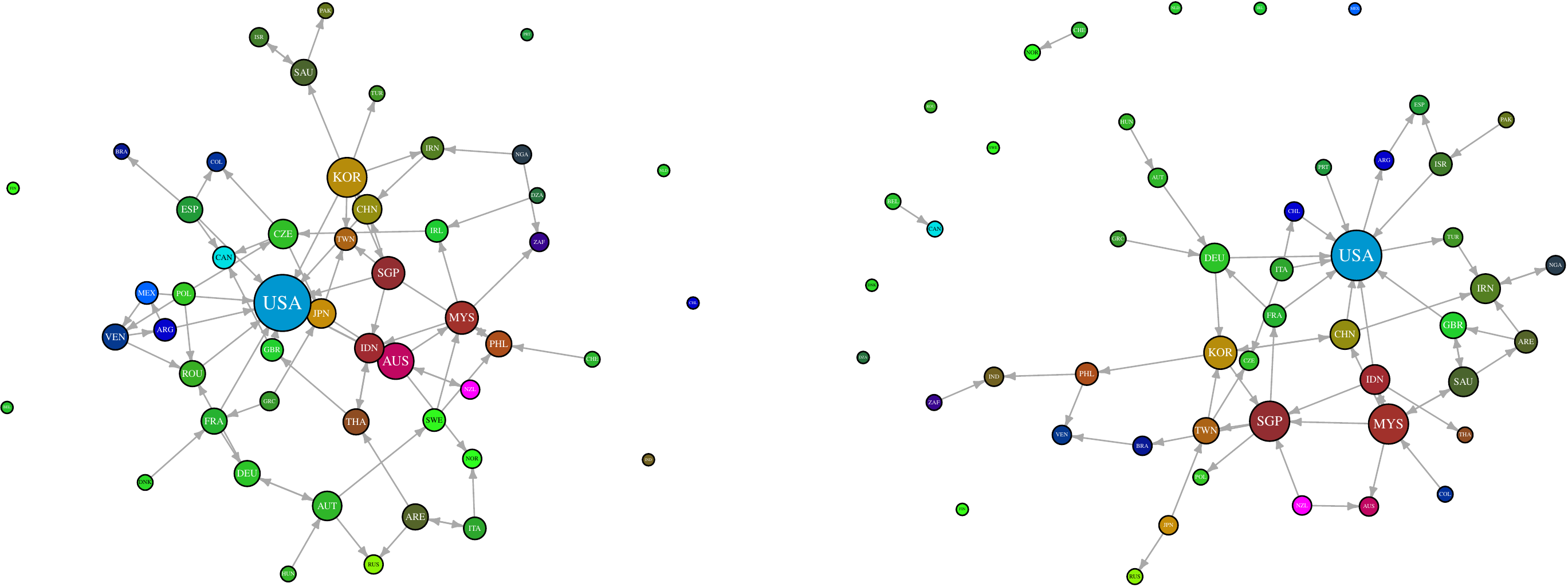}
		\label{fig:confB12}} \\
		\subfloat[][Legend]{	
			\includegraphics[width=.6\textwidth]{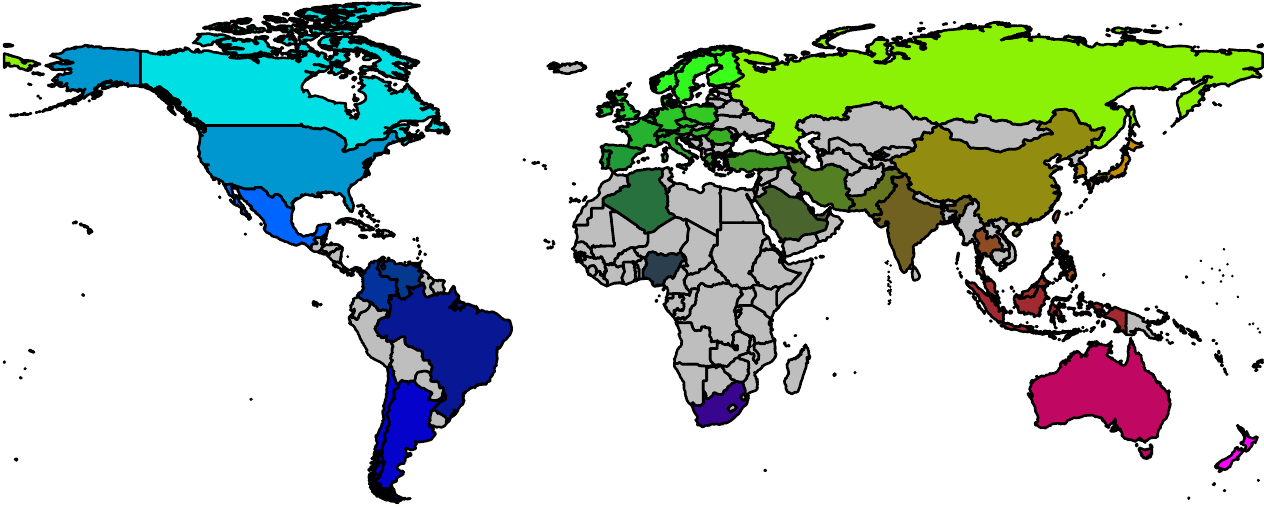}
		\label{fig:map}}		
	\end{tabular}
	\caption{$\bl B_1$, on left, and $\bl B_2$, on right. Colors are assigned to nodes by the geographic position of countries (see Figure~\ref{fig:map}) and links are drawn between nodes when the corresponding parameter estimate for that $ij$ pair is significant at a 99\% credible interval. Arrows designate the direction of the relationship. The size of nodes correspond to the number of linkages sent and received.}
	\label{fig:B12}		
\end{figure*}
\FloatBarrier

\subsection{$\bl B1_{ii'}$ \& $\bl B2_{ii'}$} 


The posterior distributions of $\bl B1_{ii'}$ and $\bl B2_{ii'}$ are summarized in Figures~\ref{fig:coopB12} for cooperation and \ref{fig:confB12} for the conflict models. Both these parameters contain $m \times m$ regression coefficients. The $lj$ coefficient in $\bl B1_{ii'}$ provides a measure of how likely $l$ is to send, for example, a verbal or material cooperative event to a third party, $k$, given that $j$ has already sent a cooperative event to $k$. Similarly, the $lj$ coefficient in $\bl B2_{ii'}$ provides a measure of how likely $l$ is to receive a cooperative event from a third party, $k$, given that $j$ has already received a cooperative event from $k$.

To visualize the posterior distributions for these parameters, we create a set of network plots in which linkages designate significant positive effects using a 99\% credible interval. Nodes are replaced with three letter country abbreviations and these pieces of text are colored by the geographic location of countries shown in Figure~\ref{fig:map}. Results using a lower threshold for significance lead to a large number of linkages being drawn and the plots becoming illegible. Additionally, very few, less than 3\%, of the $lj$ estimates were significantly negative, so we exclude visualizing those here as well to improve readability. 

Another important feature missing from our visualizations of $\bl B1_{ii'}$ and $\bl B2_{ii'}$ in Figures~\ref{fig:coopB12} and \ref{fig:confB12} is the diagonal elements. The diagonal elements, in the case of $\bl B1_{ii'}$, represent the tendency of $l$ to send a cooperative event to a third party, given that $l$ has done so in the past. Even though we do not visualize those relationships here, their effects are notable; specifically, we find that the posterior means of the diagonal elements of both $\bl B1_{ii'}$ and $\bl B2_{ii'}$ are approximately 10 times as large as the off-diagonal elements for the cooperation model. For the conflict model, the diagonal elements are about 12 times as large. This result points to a certain stickiness in how states relate to others in our sample. 

Nonetheless, there are additional findings to be gleaned from the network visualizations we present in Figures~\ref{fig:coopB12} and \ref{fig:confB12}. First, it is clear that in each of these panels we see a noticeable amount of geographic clustering. For example, in the $\bl B_1$ posterior visualization from the cooperative model, the Philippines (PHL) are likely to send cooperative events to countries that have already received cooperative verbal or material support from Malaysia (MYS). Malaysia is more likely to do the same for countries that have received cooperative interactions from Indonesia (IDN). Israel (ISR) and Turkey (TUR) also both share this type of relationship, in that a third party that has received a cooperative event from one has a higher of probability receiving that type of event from the other at a future time point as well. 

This type of geographic clustering is still apparent as we move to examining $\bl B_2$ from the  conflict model, shown in the right panel of Figure~\ref{fig:confB12}. Here we can see that Hungary (HUN) is more likely to receive conflictual events from countries that have sent those type of events to Austria (AUT). However, there are a number of relationships here that go beyond simple geographic clustering. For example, South Africa (ZAF) is significantly more likely to receive conflictual events from countries that have threatened India (IND); and Argentina is more likely to receive conflictual events from countries that have previously threatened Spain. 

\begin{figure*}[!ht]
	\centering
			\includegraphics[width=.8\textwidth]{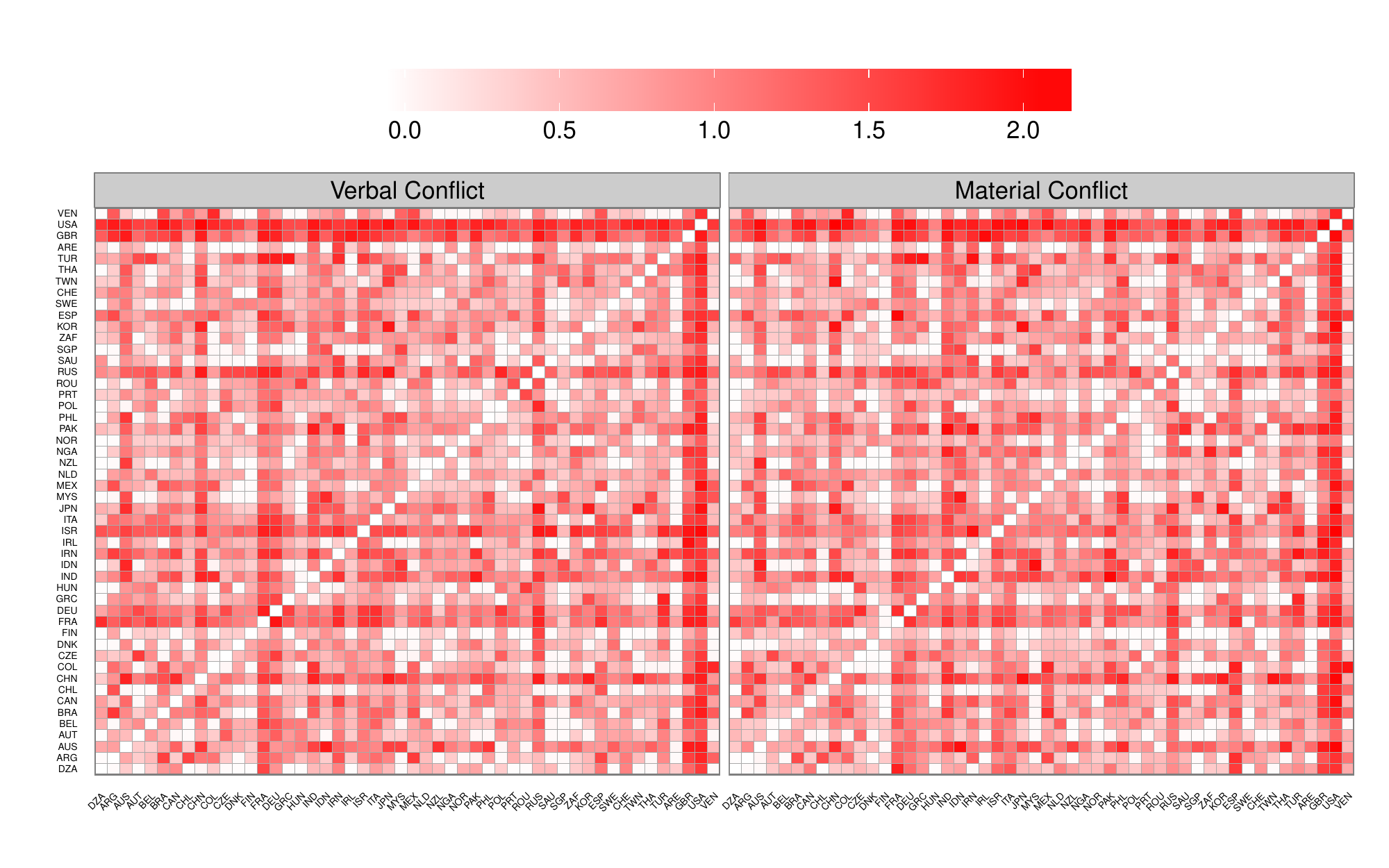}
	\caption{Root-mean-squared error values for every directed country-pair time series.}
	\label{fig:perf}
\end{figure*}

\subsection{Performance analysis}

Last, we undertake an assessment of the model's performance. To do this, we calculate the in-sample root-mean-squared error (RMSE) for the time series of every country-pair by variable. This generates an $(m \times m) - m$ number of RMSE values; we visualize the results of this analysis using the surface plot shown in Figure~\ref{fig:perf}. Here we only show results for the conflict model -- the performance of the cooperation model is similar. 

The results here highlight that the model's performance -- particularly, for pairings involving countries like France (FRA), Germany (DEU), United Kingdom (GBR), and the United States (USA) -- has meaningful room for improvement.  However, the average error of the model is not large, though it does vary substantially across dyads.  For example, the  fit of the model with regards to the largest countries is likely a result of the fact that these countries are obviously quite different from others in the international system. Thus, even with a model as detailed as that presented, there is still work to be done in contextualizing the interaction patterns. In future work we hope to address issues involving the model's performance, by allowing it to apply even more differentially to different types of nation-pairs, in a kind of bloc partitioning. 

Compared to the standard approaches, however, this novel approach generates a great deal of information about dyadic interactions that was simply lost before. Having that information also allows us to have a better way to get at first-order, as well as the second and third order interactions.

\section{Lessons Learned}

The field of International Relations has always had big data, data that are characterized as time-series cross-sections of relational data, often dyads. These data structures defeat most standard regression-based attempts to analyze their structure. In particular, some approaches such as latent variable models have been developed to handle the cross-sectional dependence and these have been applied in some of the literature, but ignored in much extant work. And some approaches have focused on the underlying structure of the time domain, but these too have been widely ignored. 

At a substantive level, we know that the world around us is highly organized in hierarchical 
ways, and that some nodes are more powerful in their interactions than others \citep{house:ward:1988,house:ward:1988a}, but that figuring this out computationally or statistically was challenging exactly because of the dependencies. At the same time, finding a way to embed the dynamical aspect of world politics in these frameworks was often relegated to a simple inclusion of time counters into regression models. But the framework herein allows this to be accomplished in a more straightforward approach that is quite familiar since it is essential a type of regression, except with network matrices stacked over time.  

The substantive findings that emerge from this suggest something both about the time dependencies and the network effects that permeate the ebb and flow of international politics.


\section*{References}
\bibliographystyle{jpr}
\bibliography{tensor, /Users/mw160/git/whistle/master, /Users/janus829/Research/WardProjects/whistle/master}
\end{document}

%% file: tensorViz.tex
\begin{tikzpicture}

	\begin{scope}[xshift=1cm, yshift=1cm]
	\node{
		\begin{tikzpicture}[scale=.5]
			 \draw[thin, black,fill=green3] (0,0) grid (4,4) rectangle (0,0) ;
		\end{tikzpicture}
	};
	\end{scope}

	\begin{scope}[xshift=.5cm, yshift=.5cm]
	\node[](green){
		\begin{tikzpicture}[scale=.5]
			 \draw[thin, black,fill=green2] (0,0) grid (4,4) rectangle (0,0) ;
		\end{tikzpicture}
	} ;
	\end{scope}

	\begin{scope}
	\node{
		\begin{tikzpicture}[scale=.5]
			 \draw[thin, black,fill=green1] (0,0) grid (4,4) rectangle (0,0) ;
		\end{tikzpicture}
	};
	\end{scope}
	
	\begin{scope}[xshift=4.5cm, yshift=1cm]
	\node{
		\begin{tikzpicture}[scale=.5]
			 \draw[thin, black,fill=blue3] (0,0) grid (4,4) rectangle (0,0) ;
		\end{tikzpicture}
	};
	\end{scope}

	\begin{scope}[xshift=4cm, yshift=.5cm]
	\node[](blue){
		\begin{tikzpicture}[scale=.5]
			 \draw[thin, black,fill=blue2] (0,0) grid (4,4) rectangle (0,0) ;
		\end{tikzpicture}
	};
	\end{scope}

	\begin{scope}[xshift=3.5cm]
	\node{
		\begin{tikzpicture}[scale=.5]
			 \draw[thin, black,fill=blue1] (0,0) grid (4,4) rectangle (0,0) ;
		\end{tikzpicture}
	};
	\end{scope}
	
	
\end{tikzpicture}